# Black phosphorus: A new bandgap tuning knob

Rafael Roldán and Andres Castellanos-Gomez

***Modern electronics rely on devices whose functionality can be adjusted by the end-user with an external 'knob'. A new tuning knob to modify the band gap of black phosphorus has been experimentally demonstrated.***

Tunability is at the hard core of electronic components. In field effect transistors (FET), for example, the charge carrier density is tuned by means of an external voltage applied to the gate electrode. The isolation of two-dimensional materials holds great promise for the fabrication of novel functional devices because their inherent reduced thickness makes them extremely sensitive to external stimuli and thus highly tunable. In particular, the recently isolated black phosphorus[1] has already demonstrated an exceptional tunability of its electronic properties by different methods including quantum confinement (sample thickness)[2], high pressure[3], mechanical strain[4,5] and chemical doping[6]. Using those approaches the band gap of black phosphorus could be varied in the range of ~2 eV to 0 eV (see Table 1). However, these tuning methods are not practical for their integration in functional nanodevices.

Now, Bingchen Deng and co-workers[7] writing in Nature Communications have found out a 'new tuning knob' to modify the electronic properties of black phosphorus. They fabricate dual-gate FETs with BP channels of different thicknesses and demonstrate bandgap tuning controlled by means of a transverse electric field. This observation can be understood as the solid-state realization of the Stark effect, commonly observed in atomic and molecular physics. In fact, while standard Stark effect leads to splitting and deviation of atomic energy levels due to presence of an external electric field, in crystalline semiconductors the application of a bias voltage can modify the energy gap due to reconstruction of the band structure. Furthermore, due to reduced electric field screening in quantum confined systems, Stark effect leads to stronger modulation of the electronic and optical properties in 2D crystals like few-layer black phosphorus flakes than in their 3D bulk counterparts.

In 2015 Qihang Liu et al. studied theoretically this problem using density functional theory (DFT) calculations and predicted that black phosphorus could exhibit a strong modulation of the bandgap in the presence of a perpendicular electric field[8]. This end is what has now been demonstrated by Deng and colleagues who, by using a four-probe dual-gate FET, have measured the conductance of BP thin films at different vertical bias voltages. The evolution of the bandgap is then obtained from the variation of the BP conductance at the charge neutrality point upon application of the perpendicular electric field and from the quantitative analysis of its temperature dependence. The electrostatic potential and the charge distribution across the BP sample is controlled by the energetic balance between the induced inter-layer capacitance and the kinetic energy terms, which strongly depend on the thickness of the sample[9]. On the other hand, it is well known the great dependence of the bandgap with the number of BP layers[2], what motivated the authors to study samples of different thicknesses (~4 nm and ~10 nm). This allowed them to explore the scaling of the bandgap tuneability with the size of the unbiased gap, in samples that present different electrostatic screening properties across them.

For thin samples of ~4 nm (which correspond to ~7 BP layers) they observe, for a moderate displacement field of ~1 V/nm, a bandgap shrinkage of ~75 meV. For 10 nm thick samples the authors find a huge increase, of about 40 times, of the minimum conductance at the charge neutrality point when the bias voltage is maximized, indicating a bandgap modification of about 300 meV, much stronger than in 4 nm samples. The observed difference is expected since, for thin samples and in the presence of moderate electric field, the inter-layer coupling dominates over the potential drop across the sample, limiting the tuneability of the bandgap. On the other hand, the effective potential difference between the top and bottom layers is stronger for thicker samples, leading to a potential drop across the sample that dominates over inter-layer hopping, enhancing the Stark effect and therefore the bandgap tuneability. The optimal tunability is found for samples with thickness comparable to that of the penetration length of the electric field in BP, ~10 nm.

Independently to the work of Deng et al.[7], Yanpeng Liu and co-workers also studied the effect of an external electric field in the bandgap value of black phosphorus[10]. Instead of measuring conductance in dual-gate FETs, Liu et al. obtained tunnelling spectra of few layer BP flakes using a low-temperature scanning tunnelling microscope (STM)[10]. Their setup consists of a few-layer BP flake deposited on $SiO_2$/Si substrate and contacted with a gold electrode. Interestingly they also measure a BP bandgap reduction from ~310 meV to ~200 meV under the application of a perpendicular electric field[10], in what can be considered an additional confirmation of tuneable Stark effect in BP.

The two methods are therefore complementary. The dual-gate device used by Deng et al.[7] has the advantage to allow for effective compensation of the doping induced by the back gate, leading to an insulating behaviour at the charge neutrality point. In any case, the modulation of the optical and electronic properties of few layer BP by external electric fields is clearly demonstrated. One of the clearest advantages of this method is reversibility, especially in comparison to chemical doping or quantum confinement tuning methods. Furthermore, electric field tuning of the bandgap can offer higher tuning speed than other tuning methods such as high pressure or strain engineering.

The development of a tuning method to dynamically adjust the bandgap of black phosphorus in a wide range can have strong implications for the implementation of optical modulators and photodetectors operating in the mid-IR. As sketched in Figure 1, unbiased BP is *transparent* for light of wavelength longer than that imposed by the bandgap (~4 μm). The application of a bias voltage reduces, in a controlled manner, the size of the semiconducting bandgap and therefore, makes the device optically active for a broad range of wavelengths, with the subsequent applicability for infrared optoelectronic devices and optical modulators or switches (see Figure 1).

Future work may aim at improving the electrostatic gating technique to achieve full closing of the bandgap. This would be of great interest because closing the bandgap in BP does not only mean to reach a semiconducting-to-semimetal transition, but it also implies a change in the topology of the material itself[11]. In short, this transition can be understood in the following manner (Figure 2). Unbiased BP is a trivial semiconductor with the direct gap placed at the Γ point of the Brillouin zone. For some critical value of the bias field, the paraboloidal valence and conduction bands merge, completely closing the band gap. Applying even higher values of the perpendicular electric field eventually leads to generation of a pair of Dirac like cones in the energy spectrum, which is accompanied by a change in the topology of the system due to emergence of $\pm\pi$ Berry phases around the Dirac points. In fact, this transition has already been experimentally observed in chemically doped BP[6] and in samples subjected to high pressure[3]. However, electrostatic gating would allow for a dynamical control of this transition and its possible use in nanodevices.

To wrap up, the independent experimental verification of the giant Stark effect in few-layer black phosphorus of Deng et al.[7] and Yanpeng Liu et al.[10] opens up a new and exciting avenue to observe new physical phenomena and to fabricate functional devices based on black phosphorus using a moderate perpendicular electric field as an external tuning knob which can be easily integrated with other existing devices architectures.

**Authors information:** Rafael Roldán[1,*] and Andres Castellanos-Gomez[1,*]

Instituto de Ciencia de Materiales de Madrid (ICMM-CSIC). 28049 Madrid. Spain.

* rroldan@icmm.csic.es, andres.castellanos@csic.es

|  | **Thickness** | **Chemical doping** | **Controlled strain** | **High pressure** | **Electrostatic gating** |
|---|---|---|---|---|---|
| **Gap modulation** | ~1.7 eV | ~0.6 eV | ~0.7 eV | ~0.3 eV | ~0.3 eV |
| **Reversibility** | Fixed band gap for a certain thickness | Not reversible | Reversible | Reversible | Reversible |
| **Prospective tunabiltiy speed** |  | Slow | Fast | Very slow | Very fast |
| **Reference** | Yang et al.[2] | Kim et al.[6] | Quereda et al.[5] | Xiang et al.[3] | Deng et al.[7] Liu et al.[10] |

Table 1. Compilation of the different methods used to manipulate the bandgap in BP.

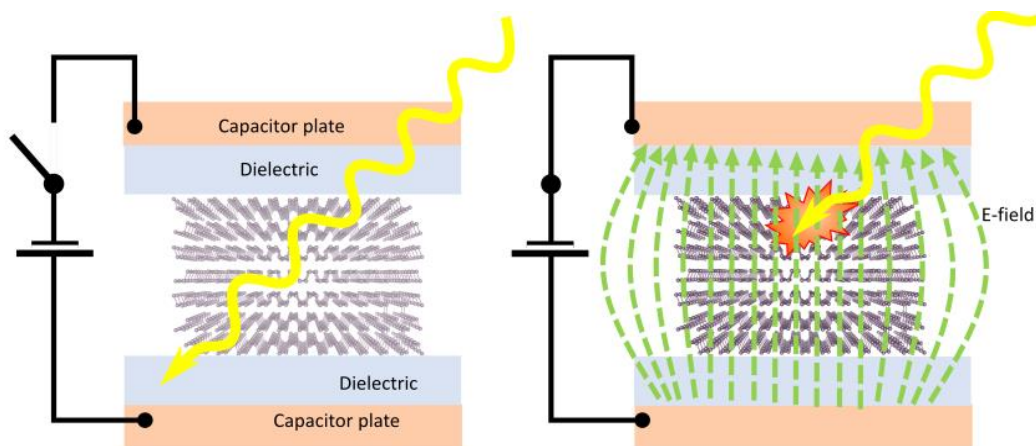

Figure 1. Sketch of the implications of the experimental discovery of a large Stark effect in BP by Deng et al.[7] and Liu et al.[10]. Pristine multilayer BP has a band gap of ~300 meV which makes it transparent to electromagnetic radiation with a wavelength longer than ~4 µm. Upon an application of a perpendicular electric field the band gap can be drastically

reduced, becoming almost metallic and thus absorbing the incoming radiation with wavelength in the ~4 µm range.

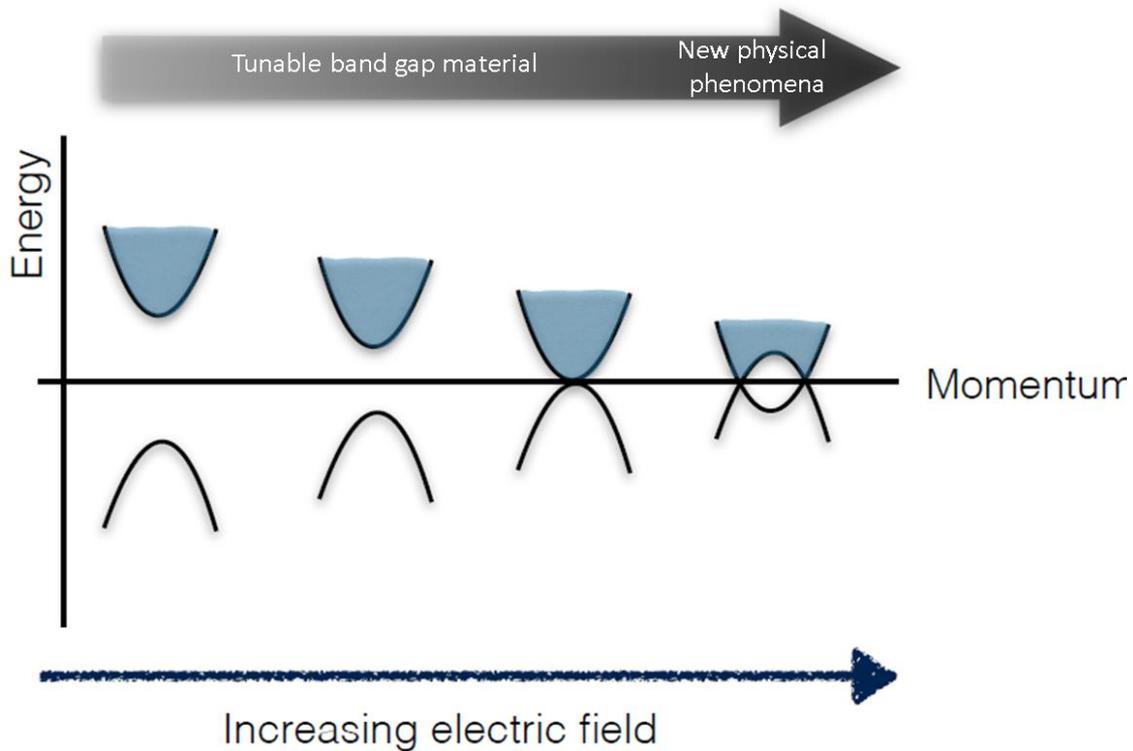

Figure 2: Sketch of the evolution of the BP band structure with applied perpendicular bias field. The bandgap is continuously reduced, making the material optically active for longer wavelengths, as denoted by the top arrow. Eventually a semiconducting-to-semimetal transition is reached, with the generation of a pair of Dirac points in the spectrum.